\documentclass[twoside]{article}
\usepackage{fleqn,espcrc2}
\usepackage{graphicx}
\usepackage[figuresright]{rotating}

\newcommand{\AmS}{{\protect\the\textfont2
  A\kern-.1667em\lower.5ex\hbox{M}\kern-.125emS}}

\title{Spectra of massive QCD Dirac Operators from Random Matrix
  Theory: all three chiral symmetry breaking patterns\thanks{presented by
  G.A. at Lattice 2000, Bangalore, India} }

\author{G. Akemann\address{Max-Planck-Institut f\"ur Kernphysik,
                           Postfach 103980, D-69029 Heidelberg, Germany}%
        \thanks{partial support by EU TMR grant {\sc erbfmrxct}97-0122}        
        and
        E. Kanzieper\address{
Hitachi Cambridge Laboratory, Madingley Road, Cambridge CB3 0HE,
United Kingdom,\\
The Abdus Salam International Centre for Theoretical Physics
P.O.B. 586, 34100 Trieste, Italy 
}} 
       
\begin{document}

\begin{abstract}
The microscopic spectral eigenvalue correlations of QCD Dirac
operators in the presence of dynamical fermions are calculated 
within the framework of Random Matrix Theory (RMT).
Our approach treats the low--energy correlation functions 
of all three chiral symmetry breaking patterns (labeled
by the Dyson index $\beta=1,2$ and 4) on the same footing, 
offering a unifying description of massive QCD Dirac spectra. RMT
universality is explicitly proven for all three symmetry classes
and the results are compared to the available lattice
data for $\beta=4$.
\end{abstract}

\maketitle

\section{Introduction}

Random Matrix Theory (RMT) has turned out to be a very fruitful
tool in studying the phenomenon of chiral symmetry breaking in 
low--energy QCD \cite{VW}. 
First proposed as a purely phenomenological approach, 
it has recently been put onto firm field theoretic grounds after the 
analytic RMT predictions have been reproduced within the framework 
of finite--volume partition functions and partially quenched 
chiral perturbation theory using supersymmetry \cite{DOTV} and
replica  \cite{DS} techniques.

Similarly to previously studied sum rules \cite{LS}, RMT solutions 
for spectral statistics serve as a more detailed test of QCD
lattice data, in particular for a given sector of topological charge
$\nu$ (which counts the number of zero modes of the QCD Dirac
operator).  
This has become
possible due to recent developments in lattice gauge theory. 
Namely, the Ginsparg-Wilson relation was shown to provide an exact chiral
symmetry on the lattice together with a well defined topological
charge $\nu$ \cite{N} (and these proceedings). 
The classification of the three different $\chi$SB patterns according to
gauge group and representation \cite{V94} 
has been confronted to lattice data and good agreement 
has been found for the spectral density and 
distribution of the smallest Dirac eigenvalue for {\it massless} 
lattice data of all $\chi {\rm SB}$ patterns and for different values of
$\nu$ (see e.g. 
\cite{EHKN}, Fig. 2). 

It has to be mentioned that the direct  field theoretic calculation of spectral
correlators is much more cumbersome as compared to RMT. 
The former approach has not led to any explicit analytic results beyond the 
spectral density due to enormous increase of 
dimensionality of auxiliary (supermatrix) fields involved. So far, only 
spectral densities for massless flavors (in all three 
$\chi$SB patterns) and for one single massive flavor (in
$SU(N_c\!\geq\!3)$ in the fundamental representation) have been
derived \cite{DOTV}. In contrast, the classic RMT technique is free 
of the above technical complications and allows computing the higher 
order correlation functions with the same ease
(see Ref. \cite{MGW} for comparison with lattice data).
This advantage of a RMT 
description becomes even more significant for dynamical fermions.
The correlation functions with an arbitrary number of massive flavors
$N_f$ have been calculated 
for fundamental $SU(N_c\!\geq\!3)$ \cite{JNZ,DNW} $(\beta\!=\!2)$ 
and recently for the two remaining $\chi$SB
patterns \cite{AK,NN} $(\beta\!=\!1,4)$.

In the present communication, we 
report on our results \cite{AK} 
relevant for
gauge groups
$SU(2)$ in the fundamental ($\beta=1$) and $SU(N_{c})$ 
in the adjoint representation ($\beta=4$). Our predictions are 
compared to $SU(2)$ lattice data \cite{BMW} for dynamical staggered 
fermions with 4 degenerate flavors.

\section{RMT results for massive flavors}

Let us briefly recall the connection between
RMT and low--energy QCD. The Dirac operator spectrum at the origin is
related to the chiral condensate $\Sigma$, the order parameter of $\chi$SB,
through the Banks-Casher relation \cite{BC}
$\Sigma=\lim_{V\to\infty}\pi\rho(0)/V$, where $V$ is the Euclidean
space-time volume. Here, the spectral density of the Dirac operator is
given by the average
$\rho(\lambda)\!=\!\langle\sum_k\delta(\lambda-\lambda_k)\rangle$ over 
all gauge field configurations and the $\lambda_k$ are
the Dirac operator eigenvalues. In the limit
$\Lambda^{-1}\!\ll\! V\!\ll\! m_\pi^{-1}$ \cite{LS}, where $m_\pi$ is the 
pion mass and $\Lambda$ is the scale of the lightest non-Goldstone
particle, the QCD partition function is dominated by zero momentum modes of the
Goldstone fields and hence collapses into a simple group
integral \cite{LS}. As a result, the partition function only depends 
on the global symmetries of the QCD Dirac operator and contains just
the rescaled quark masses $\mu_f\!=\!m_fV\Sigma$ as parameters.
In a sector with
fixed topological charge $\nu$, it coincides with the corresponding RMT
partition function once the space-time volume $V$ is identified 
with the size $n$ of the corresponding random matrix.
Analogously, all correlation functions can also be computed from RMT provided
the matrix eigenvalues are appropriately rescaled,  
$\xi_k\!=\!\lambda_kV\Sigma\!=\!\lambda_k n\pi\rho(0)$,
where $V\to\infty$ or $n\to\infty$
is taken;  $\rho(0)$ denotes the RMT spectral density.
This provides us with a parameter free prediction. 

The joint probability density function of chiral RMT
associated with $N_f$ massive 
quarks in the sector of topological charge $\nu$ is defined as 
\begin{eqnarray}
\label{jpdf}
&&P_n^{(N_f,\nu ,\beta )}(\lambda_1,\ldots,\lambda_n ) = 
\frac{1} {Z_n^{(N_f,\nu ,\beta )}(\{m\})} \\
&&\times \left| \Delta _n\left( \{ \lambda\} \right) \right|^\beta
\prod_{i=1}^n [ w_{\beta,\nu }(\lambda _i)\!
\prod_{f=1}^{N_f}m_f^\nu(\lambda _i+m_f^2)]. \nonumber
\end{eqnarray}
Here, $\beta\!=\!1,2$ and $4$ labels the symmetry of the 
matrix ensemble to be 
orthogonal ($\beta\!=\!1$), unitary ($\beta\!=\!2$) or symplectic
($\beta\!=\!4$)  
in correspondence with the three $\chi$SB patterns
\cite{V94}.
The partition function appearing in the normalization is obtained by 
integrating over all
eigenvalues $\lambda_k$. 
The $k$-point
correlation function is determined by integrating over $n-k$
eigenvalues only:
\begin{eqnarray}
\label{k-pt}
&&R_{n,k}^{(N_f,\nu ,\beta )}(\lambda _1,\ldots,\lambda _k) \ =  
\frac{n!}{(n-k)!} \\
&&\times \int_0^{+\infty }\!\!d\lambda _{k+1}\ldots d\lambda _n\ 
P_n^{(N_f,\nu,\beta
  )}(\lambda_1,\ldots,\lambda_n). 
\nonumber
\end{eqnarray}
Here, 
$\Delta _n\left( \{\lambda\} \right)=\prod_{i<j}^n(\lambda_i-\lambda_j)$ 
is the Vandermonde determinant and the weight function
is given by
\begin{eqnarray}
\label{w}
&&w_{\beta ,\nu }( \lambda ) =\lambda ^{\frac \beta
2\nu +\frac \beta 2-1}e^{- \beta V\left( \lambda \right)},
\end{eqnarray} 
where $V(\lambda)$ is a finite-polynomial confinement potential
whose form is not fixed apriori. Although the simplest choice
$V(\lambda)=\Sigma^2\lambda$ defining the Gaussian ensemble leads to
significant mathematical simplifications it cannot be 
derived from the QCD Lagrangian. 
It is therefore crucial to show
that the RMT results for
the rescaled $k$-point correlation functions, Eq. (\ref{microk-pt}), 
are universal and do not depend\footnote{The only condition is that
  the macroscopic RMT spectral density has to obey $\rho(0)\neq0$.} 
on this choice for $V(\lambda)$. 

In the following we present a unified way to explicitly calculate, 
and prove, the RMT universality of massive spectral correlators 
for all three $\chi{\rm SB}$ patterns, $\beta\!=\!1,2$ and $4$. Our 
strategy is to express the {\it massive} spectral correlators in terms 
of the known 
{\it massless} ones; the latter have already been shown to be universal 
\cite{ADMN}. To proceed, we assume that the massive fermions are $\beta$--fold 
degenerate. With the help of the identity
\begin{eqnarray}
&&\frac{\Delta _{n+N_f}(\{\lambda \},\{-m^2\})}{\Delta_{N_f}(\{-m^2\})}
\nonumber\\
&&=\ \Delta_n(\{\lambda \})
\prod_{i=1}^n\prod_{f=1}^{N_f}(\lambda_i+m_f^2)\ , 
\label{delta}
\end{eqnarray}
the joint probability density $P^{(\beta 
N_{f},\nu,\beta)}_{n}$ associated with the $\beta$--fold degenerate 
massive fermions of total amount $\beta N_{f}$ can be rewritten 
through the
massless joint probability density
$P_{n+N_f}^{(0,\nu ,\beta )}$ with $n$ positive $\{\lambda_i\}$ and $N_f$ 
negative $\{-m_f^2\}$ eigenvalues. This 
leads us to the 
remarkable identity
\cite{AK} which holds in all generality for finite $n$:
\begin{eqnarray}
\label{master-n}
&&R_{n,k}^{(\beta N_f,\nu ,\beta )}(\lambda _1,\ldots,\lambda _k) \\
&&=\frac{R_{n+N_f,k+N_f}^{(0,\nu ,\beta )}(\lambda _1,\ldots,\lambda_k,
-m_1^2,\ldots,-m_{N_f}^2)}{R_{n+N_f,N_f}^{(0,\nu ,\beta)}
(-m_1^2,\ldots,-m_{N_f}^2)} \nonumber
\end{eqnarray}
In order to compare with QCD
we have to perform the microscopic
large-$n$ limit as mentioned above. 
It reduces to evaluating
the rescaled (or unfolded) correlators 
\begin{eqnarray}
&&\rho_S^{(N_f,\nu,\beta)}(\xi_1,\dots,\xi_k) \ = \ 
\lim_{n\to\infty} \prod_{i=1}^k
\left(\frac{2|\xi_i|}{n\pi\rho(0)}\right) 
\nonumber\\
&&\times\ 
R_{n,k}^{(N_f,\nu,\beta)}\!\left( \frac{\xi_1^2}{(n\pi\rho(0))^2},
  \dots,\frac{\xi_k^2}{(n\pi\rho(0))^2}
\right)
\label{microk-pt}
\end{eqnarray}
with a similar rescaling of the masses. 
Here, we have switched from positive
to real Dirac operator eigenvalues. 
It is easy to see that Eqs. 
(\ref{master-n}) and (\ref{microk-pt}) result in
the following expression for the microscopic $k$-point correlation function 
with $\beta N_f$ masses:
\begin{eqnarray}
\label{master}
&&\rho_{S}^{(\beta N_f,\nu ,\beta)}(\xi_1,\ldots,\xi_k) \nonumber\\
&&=\ \frac{\rho_{S}^{(0,\nu ,\beta )}(\xi_1,\ldots,\xi_k,
i\mu_1,\ldots,i\mu_{N_f})}{\rho_{S}^{(0,\nu ,\beta)}
(i\mu_1,\ldots,i\mu_{N_f})}\ .
\end{eqnarray}
Here, $\rho_{S}^{(0,\nu,\beta)}$ is the {\it massless} correlation 
function which is entirely known \cite{Mehta} in terms of determinants
$(\beta=2)$ or quaternion determinants $(\beta=1,4)$. 
Since the universality of massless correlation functions has already 
been firmly established, the universality of the 
massive ones 
automatically follows. 
Alternative representations 
of massive correlation functions 
were derived in Ref. \cite{NN} using Gaussian ensembles.
There, the mass degeneracy for $\beta\!=\!4$ 
is partially lifted to be two-fold.

In the simplest situation of the spectral density with $\beta$ 
degenerate massive fermions, Eq. (\ref{master}) reduces to
\begin{equation}
\label{k=1massive}
\rho_{S}^{(\beta,\nu ,\beta)}(\xi)
= 
\rho_{S}^{(0,\nu ,\beta)}(\xi)+
\frac{\rho_{S}^{(0,\nu ,\beta )}(\xi,i\mu)_{\rm conn}}
{\rho_{S}^{(0,\nu,\beta)}(i\mu)}\ . 
\end{equation}
The full density with $\beta$ dynamical flavors is thus given by the quenched 
density $\rho_{S}^{(0,\nu,\beta)}$ and the mass dependent correction 
term expressed through the connected part of the massless two-point 
correlation function $\rho_{S}^{(0,\nu,\beta)}(\xi,i\mu)_{\rm conn}$.

The application of Eq. (\ref{k=1massive}) to the symmetry class $\beta=1$ was 
discussed in Ref. \cite{AK} and we will not consider it in what follows. 
However, it is instructive to consider the simplest example,
the symmetry class $\beta=2$. The 
connected part of the two-point correlation function is proportional to
the square of the unitary kernel
\begin{equation}
\label{ker-2}
K_\alpha (\xi,\eta) = 
\frac{\xi J_{\alpha+1}(\xi)J_\alpha(\eta)-\eta J_{\alpha+1}(\eta)
J_\alpha(\xi)}{2(\xi^2-\eta^2)} .
\end{equation}
Combining Eqs. (\ref{k=1massive}) and (\ref{ker-2}), we easily
arrive at the result of 
Ref. \cite{DNW} for the microscopic density with two degenerate flavors,
\begin{eqnarray}
\label{k=1massive2}
\rho_{S}^{(2,\nu ,2)}(\xi)
=
\rho_{S}^{(0,\nu ,2)}(\xi) -2|\xi|
\frac{K_\nu (\xi,i\mu)^2}{K_\nu (i\mu,i\mu)}.
\end{eqnarray}
Interestingly, the result of Ref. \cite{DNW} for non-degenerate masses
$\mu_1$ and
$\mu_2$ can be put into the same form
\begin{eqnarray}
\rho_{S}^{(2,\nu,2)}(\xi)=
\rho_{S}^{(0,\nu,2)}(\xi)-2|\xi|
\frac{K_\nu (\xi,i\mu_1)K_\nu (\xi,i\mu_2)}{K_\nu (i\mu_1,i\mu_2)} \nonumber
\end{eqnarray}
of the quenched density plus a mass-dependent  correction.

At $\beta\!=\!4$, the microscopic density for 
four degenerate massive fermions is given by Eq. (\ref{k=1massive})
with the massless microscopic density \cite{NF}
\begin{eqnarray}
    \label{dd}
&&\rho_{S}^{(0,\nu ,4)}(\xi)=2|\xi|\ \big[
2K_{2\nu+1}(2\xi,2\xi) \nonumber\\
&&-\ \frac{J_{2\nu}(2\xi)}{4\xi}
 \int_0^{2\xi}\!dtJ_{2\nu+2}(t)\big]\ ,
\end{eqnarray}
and the connected part of the massless
two-point correlation function \cite{MGW}
\begin{eqnarray}
\label{k=1massive4}
&&\rho_{S}^{(0,\nu ,4)}(\xi,\eta)_{\rm conn} \\
&&=-f(\xi,\eta)\partial_\xi\partial_\eta f(\xi,\eta)
+\partial_\xi f(\xi,\eta)\partial_\eta f(\xi,\eta)\ , \nonumber
\end{eqnarray}
where 
\begin{eqnarray}
f(\xi,\eta)=\frac{\eta}{2}\int_0^{2\xi}\!\!dt K_{2\nu} (t,2\eta)
-
\frac{\xi}{2}\int_0^{2\eta}\!\!dt K_{2\nu} (2\xi,t). \nonumber 
\end{eqnarray}
Performing the analytic continuation from $J$-Bessel to $I$-Bessel 
functions 
completes the solution of Eq. (\ref{k=1massive}).

\begin{figure}[-t]
\centerline{
\includegraphics[width=10pc,angle=90]{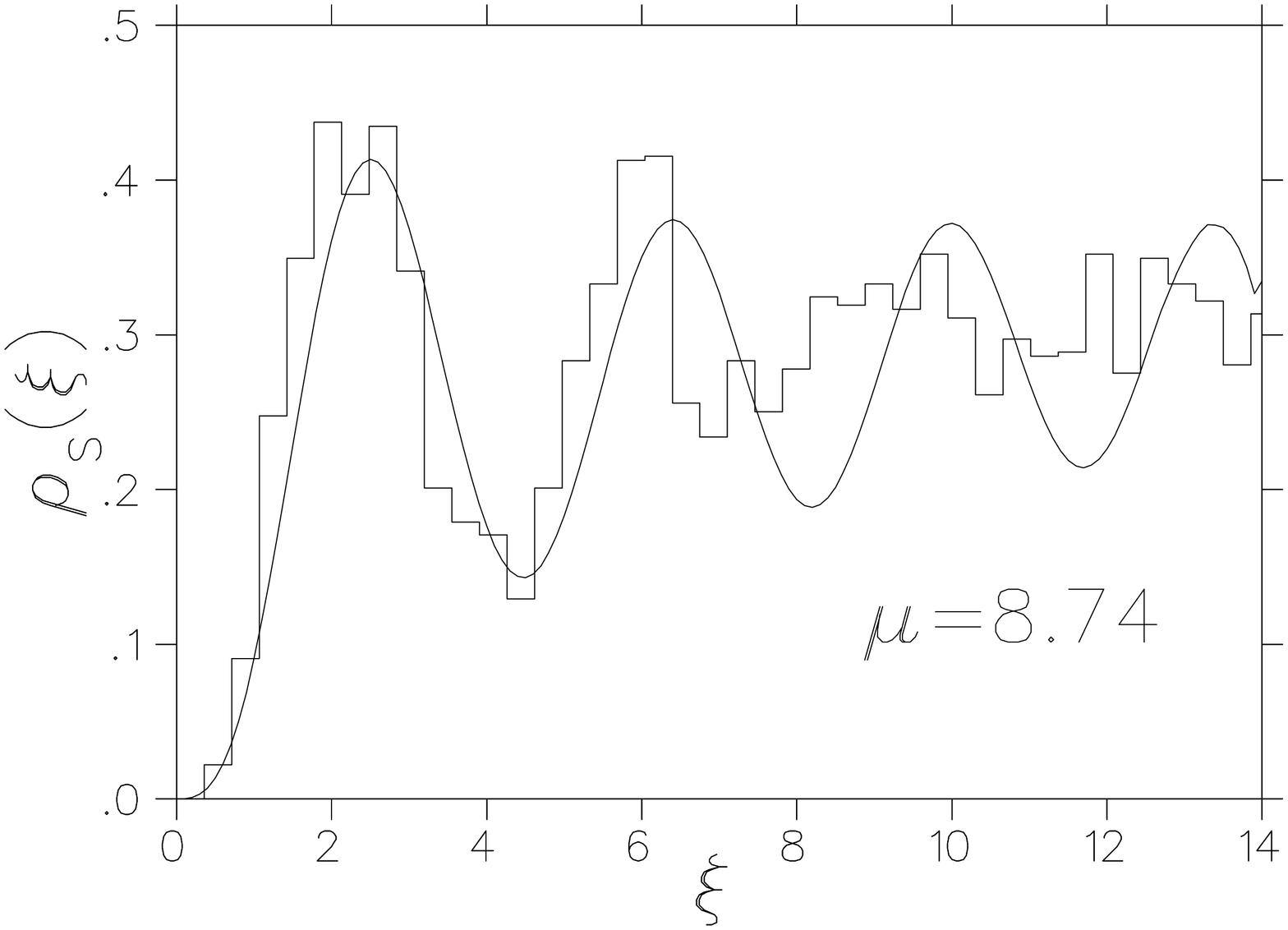}} 
\centerline{\includegraphics[width=10pc,angle=90]{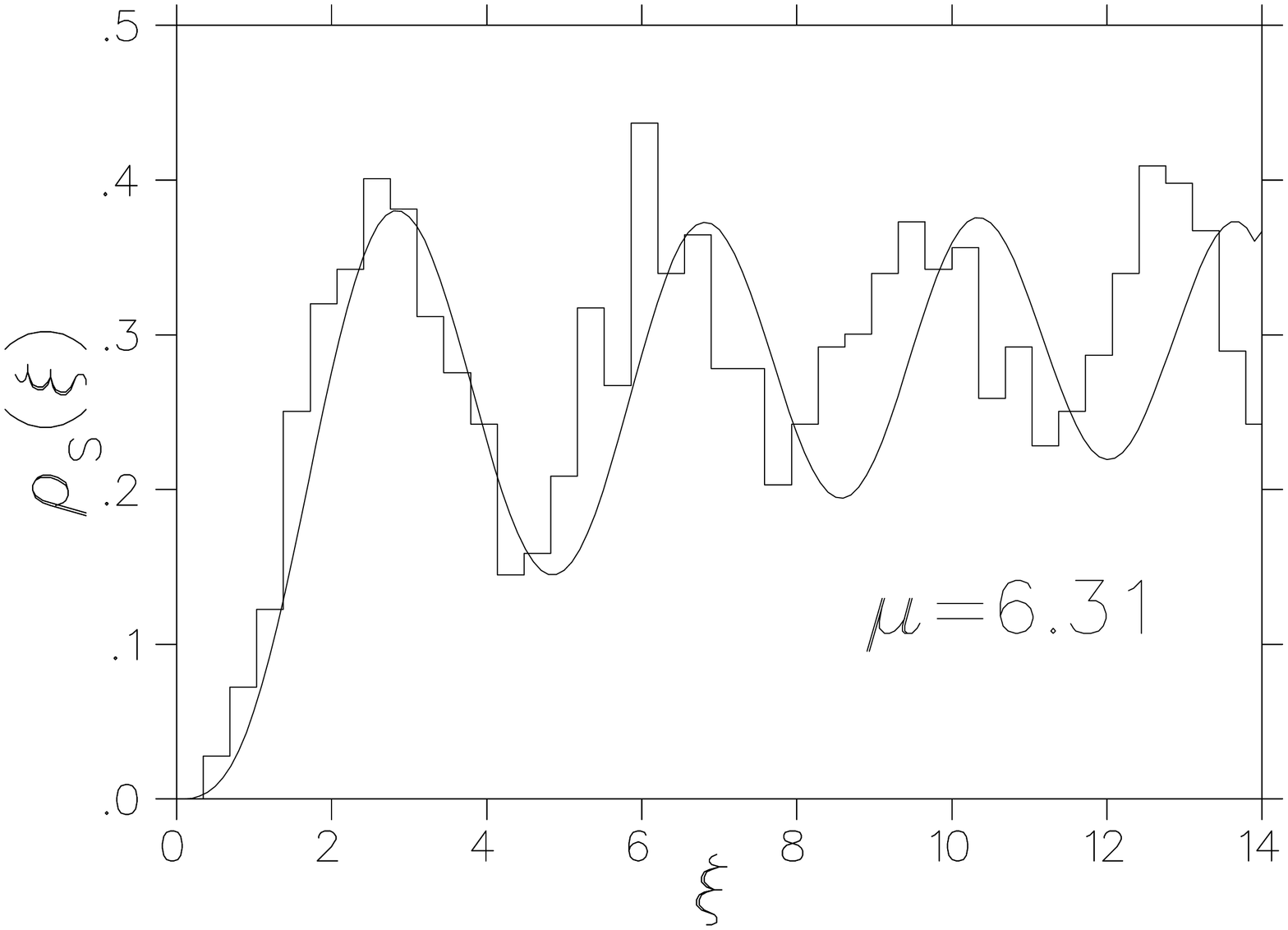}}
\centerline{\includegraphics[width=10pc,angle=90]{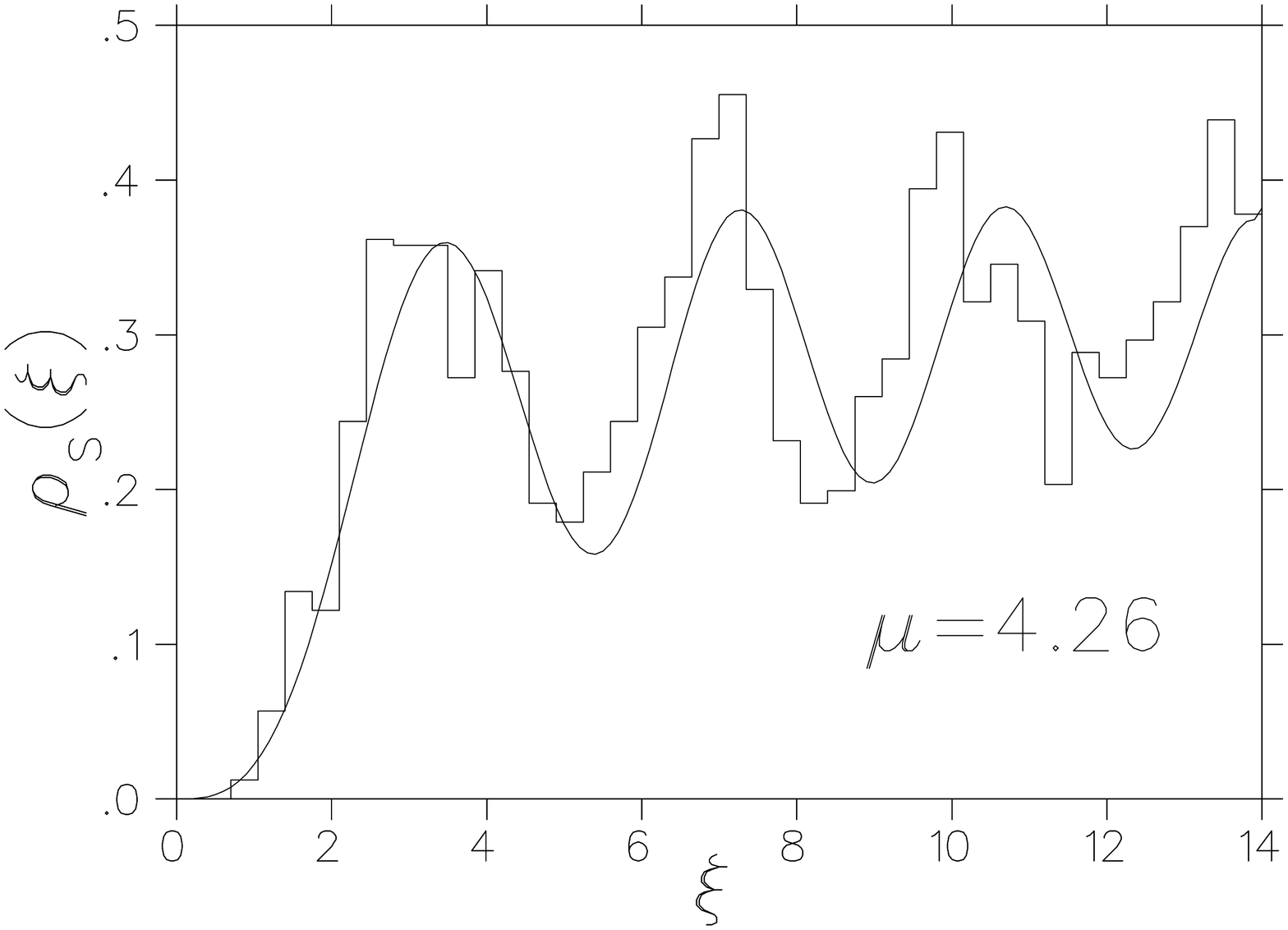}}
\vspace*{-10mm}
\caption{
The microscopic density $\rho_S(\xi)$ plotted
against lattice data for different values of $\mu$.}
\label{fig1}
\vspace*{-5mm}
\end{figure}
In Fig. 1 the microscopic massive density $\rho_{S}^{(4,\nu,4)}$ 
described by Eqs. (\ref{k=1massive}), (\ref{dd}) and (\ref{k=1massive4})
is plotted for $\nu\!=\!0$ versus the lattice data 
of Ref. \cite{BMW} 
with gauge group $SU(2)$ in the fundamental
representation. 
Because of using staggered fermions 
symmetry class $\beta=4$ applies.
A reasonable agreement between our 
parameter--free theoretical prediction and the lattice data is 
observed.
The chiral
condensate has been obtained from the Banks-Casher
relation. A fit to the best value of $\Sigma$ could improve the
systematic shift for higher values of $\xi$ due to
finite-size effect and statistics as is discussed in 
Ref. \cite{BMW}.

\end{document}